\pdfoutput=1
\documentclass[12pt,preprint]{aastex}
\usepackage[latin9]{inputenc}
\usepackage{array}
\usepackage{amsbsy}
\usepackage{amstext}
\usepackage{graphicx}

\makeatletter

\providecommand{\tabularnewline}{\\}

\makeatother

\begin{document}

\title{Dependence of stellar magnetic activity cycles on rotational period
in nonlinear solar-type dynamo}

\author{V.~V.~Pipin$^{1}$, A.~G.~Kosovichev$^{2}$}

\affil{$^{1}$Institute of Solar-Terrestrial Physics, Russian Academy of
Sciences, Irkutsk, 664033, Russia}

\affil{$^{2}$ New Jersey Institute of Technology, Newark, NJ 07102, USA}
\begin{abstract}
We study turbulent generation of large-scale magnetic fields using
nonlinear dynamo models for solar-type stars in the range of rotational
periods from 14 to 30 days. Our models take into account non-linear
effects of dynamical quenching of magnetic helicity, and escape of
magnetic field from the dynamo region due to magnetic buoyancy. The
results show that the observed correlation between the period of rotation
and the duration of activity cycles can be explained in the framework
of a distributed dynamo model with a dynamical magnetic feedback acting
on the turbulent generation either from magnetic buoyancy or magnetic
helicity. We discuss implications of our findings for the understanding
of dynamo processes operating in solar-like stars.
\end{abstract}


\section{Introduction}

The solar-type cyclic magnetic activity is often observed among main-sequence
stars with external convective envelopes (e.g., \citealp{Baliunas1995,Hall07AJ}).
After \citet{P55}, it is widely believed that the magnetic activity
on solar-like stars results from large-scale dynamo processes driven
by turbulent convection and rotation. Interpretation of the stellar
magnetic activity is rather complicated because of nonlinear dynamo
effects that needs to be taken into account \citep{1984ApJ...287..769N}.
Moreover, even in the case of the solar dynamo many details are poorly
known \citep{chrev05}. In particular, the origin of the large-scale
poloidal magnetic field of the Sun (the component of the field which
lies in meridional planes) is not well understood.

One of the interesting question is how observations of stellar magnetic
activity can help us in understanding the key processes of the solar
dynamo and vice verse \citep{Brunrev14}. The dependence of magnetic
activity cycles on the period of rotation is of particular interest.
Observations show that the stellar magnetic activity grows almost
linearly with the increasing rotation rate \citep{vid14MNR}, but
the dynamo period decreases \citep{1984ApJ...287..769N}. Thus, the
solar-like stars show an anti-correlation between the dynamo period
and the amplitude of magnetic activity. It is interesting that this
relationship can be deduced from the solar activity data \citep{vetal86},
as well. A comparative study of this relation for solar and stellar
cycles was published by \citet{soon94}. Further studies revealed
that this correlation is not unique, and that several branches corresponding
to different magnetic activity levels can be identified \citep{1999ApJ_sa_br,2007ApJ_bohm,2011IAUS_saar}.

Theoretically, the observed correlation between the amplitude and
period of the dynamo cycle is expected in a kinematic regime \citep{1984ApJ...287..769N}.
However, results of such linear analysis can not be applied to the
nonlinear dynamo. Theoretical arguments of \citet{1984ApJ...287..769N},
and, also, \citet{tob98} and \citet{oss97} had shown that dynamo
saturation mechanisms affect the correlation. Also, the dynamo scenario
type is important in this context. The concept of flux-transport dynamo
has been popular in the solar community \citep{chrev05}. However,
the stellar dynamo models which were constructed following this idea
show a growth of the cycle period with an increase of the rotation
rate, (e.g., see, \citealt{brun10AA}, and \citealt{2014ApJ...791...59K}).
This contradicts to the observations. The solution of this issue is
not clear yet. \citet{brun10AA} found that in the framework of the
flux-transport models this issue can be resolved by assuming certain
multiple-cell patterns of the meridional circulation. However, it
is unclear how such circulation patterns are compatible with the angular
momentum balance in stars. \citet{pi15M} argued that the problem
with the flux-transport models can be related to the direction of
the dynamo wave propagation in the convective zone. In the flux-transport
models the dynamo wave propagates inward from high to low latitudes
towards the regions with low magnetic diffusivity. This causes an
increase of the dynamo period when the large-scale toroidal field
concentrates stronger at the bottom of the convection zone. Such effect
has been found in 3D dynamo simulations by Guerrero et al (2016).
In the distributed dynamo models \citep{bran05,pk11apjl} the dynamo
wave propagates outwards from high to low latitudes. This type of
dynamo agrees with observations in case both of the solar \citep{pk11,pipea2012AA}
and stellar cycles \citep{pi15M}.

In this paper we study the influence of nonlinear dynamo saturation
mechanisms on basic properties of the magnetic activity cycles in
solar-like stars with different rotation rates. We restrict ourselves
by slowly rotating stars with the rotational periods from 14 to 30
days. In these cases the magnetic feedback on the differential rotation
is not very strong compared to the saturation caused by magnetic buoyancy
and magnetic helicity. From the results of \citet{pi15M} it follows
that the strongest variations of the latitudinal rotational shear
due to the dynamo-generated magnetic activity are about 1\% of the
mean value for the case of a solar analog rotating with the period
of 14 days. This agrees with the observational findings of \citet{2011IAUS_saar},
as well. The impact of such variations on the dynamo processes is
not as strong as of the others nonlinear effects. Thus, we neglect
variations of the differential rotation in this study.

\section{Basic equations}

\subsection{Dynamo model}

Following \citet{KR80}, we explore the evolution of the induction
vector of the mean magnetic field, $\overline{\mathbf{B}}$, in a
highly conductive turbulent media with the mean flow velocity, $\overline{\mathbf{U}}$,
and the mean electromotive force, $\boldsymbol{\mathcal{E}}=\overline{\mathbf{u}\times\mathbf{b}}$
(hereafter MEMF), where $\mathbf{u}$ and $\mathbf{b}$ are small-scale
turbulent fluctuations of the flow velocity and magnetic field: 
\begin{equation}
\frac{\partial\overline{\mathbf{B}}}{\partial t}=\boldsymbol{\nabla}\times\left(\boldsymbol{\mathcal{E}}+\overline{\mathbf{U}}\times\overline{\mathbf{B}}\right)\text{.}\label{eq:dyn}
\end{equation}
We employ a decomposition of the axisymmetric field into a sum of
the azimuthal (toroidal) and poloidal (meridional plane) components:
\[
\overline{\mathbf{B}}=\mathbf{e}_{\phi}B+\nabla\times\frac{A\mathbf{e}_{\phi}}{r\sin\theta},
\]
where $\mathbf{e}_{\phi}$ is the unit vector in the azimuthal direction,
$\theta$ is the polar angle, and $A\mathbf{e}_{\phi}$ is the vector-potential
of the large-scale poloidal magnetic field. The mean electromotive
force, $\boldsymbol{\mathcal{E}}$, is expressed as follows: 
\begin{equation}
\mathcal{E}_{i}=\left(\alpha_{ij}+\gamma_{ij}\right)\overline{B}_{j}-\eta_{ijk}\nabla_{j}\overline{B}_{k}.\label{eq:EMF-1}
\end{equation}
The tensor, $\alpha_{ij}$, models the generation of the magnetic
field by the $\alpha$ effect; the anti-symmetric tensor, $\gamma_{ij}$,
controls pumping of the large-scale magnetic fields in the turbulent
media; the tensor, $\eta_{ijk}$, governs the turbulent diffusion
and takes into account the generation of the magnetic fields by the
$\Omega\times J$ effect \citep{rad69}. We take into account the
effect of rotation and magnetic field on the mean-electromotive force.
Here, we study of the dynamo saturation due to the nonlinear $\alpha$
effect and magnetic buoyancy.

The $\alpha$ effect takes into account the kinetic and magnetic helicities
in the following form: 
\begin{eqnarray}
\alpha_{ij} & = & C_{\alpha}\sin^{2}\theta\psi_{\alpha}(\beta)\alpha_{ij}^{(H)}\eta_{T}+\alpha_{ij}^{(M)}\frac{\overline{\chi}\tau_{c}}{4\pi\overline{\rho}\ell^{2}}\label{alp2d-1}
\end{eqnarray}
where $C_{\alpha}$ is a free parameter, the $\alpha_{ij}^{(H)}$
and $\alpha_{ij}^{(M)}$ express the kinetic and magnetic helicity
parts of the $\alpha$ effect, respectively; $\overline{\chi}$ is
the small-scale magnetic helicity, $\ell$ is the typical length scale
of the turbulence, and $\eta_{T}$ is the turbulent diffusivity. Tensors
$\alpha_{ij}^{(H)}$ and $\alpha_{ij}^{(M)}$ depend on the Coriolis
number, $\Omega^{*}=4\pi{\displaystyle \frac{\tau_{c}}{P_{rot}}}$,
where $P_{rot}$ is the rotational period, $\tau_{c}$ is the convective
turnover time. The reader can find a detailed description of these
parameters in the paper of \citet{pi15M}. The magnetic quenching
function of the hydrodynamical part of the $\alpha$ effect is defined
as:
\begin{equation}
\psi_{\alpha}=\frac{5}{128\beta^{4}}\left(16\beta^{2}-3-3\left(4\beta^{2}-1\right)\frac{\arctan\left(2\beta\right)}{2\beta}\right),\label{eq:static}
\end{equation}
where $\beta=\left|\mathbf{B}\right|/\sqrt{4\pi\overline{\rho}u'^{2}}$,
$u'$ is the RMS of the convective velocity. This is so-called ``algebraic''
quenching describes the ``instantaneous'' magnetic feedback on the
dynamo generation. It is assumed that the large-scale magnetic field
varies much slower than the typical convective time-scale. In addition,
there is a so-called ``dynamical'' quenching due to the magnetic
helicity conservation \citep{pouquet-al:1975b}. Similarly to \citet{kleruz82},
\citet{bl-br2003} and \citet{kle2003AA} (also, see, \citealt{kitiash09,sok2013,blsbr13MN}),
we model this effect though the second term of Eq.(\ref{alp2d-1})
and the equation which governs the evolution of the helicity density
of the fluctuating part of magnetic field, $\overline{\chi}$, \citep{hub-br12,pip2013ApJ}:
\begin{equation}
\frac{\partial\overline{\chi}^{(tot)}}{\partial t}=-\frac{\overline{\chi}}{R_{m}\tau_{c}}-2\eta\overline{\mathbf{B}}\cdot\mathbf{\overline{J}}-\boldsymbol{\nabla\cdot}\boldsymbol{\boldsymbol{\mathcal{F}}}^{\chi},\label{eq:helcon-1}
\end{equation}
where $\overline{\chi}^{(tot)}=\overline{\chi}+\overline{\mathbf{A}}\cdot\overline{\mathbf{B}}$
is the total magnetic helicity density, and the $\boldsymbol{\boldsymbol{\mathcal{F}}}^{\chi}$
is the diffusive flux of the magnetic helicity, $R_{m}$ is the magnetic
Reynolds number, and $\eta$ is the microscopic magnetic diffusivity.

The turbulent mean-field pumping contains a sum of contributions due
to the mean density gradient \citep{kit:1991}, $\gamma_{ij}^{(\rho)}$,
the diamagnetic pumping \citep{1987SvAL...13..338K}, $\gamma_{ij}^{(\eta)}$,
the mean-field magnetic buoyancy \citep{kp93}, $\gamma_{ij}^{(b)}$,
and due to effects of the large-scale shear \citep{2013GApFD.107..185P},
$\gamma_{ij}^{(H)}$:

\begin{equation}
\gamma_{ij}=\gamma_{ij}^{(\rho)}+\gamma_{ij}^{(\eta)}+\gamma_{ij}^{(b)}+\gamma_{ij}^{(H)}.\label{eq:pump}
\end{equation}
In our study the most essential pumping is related to the mean-field
magnetic buoyancy:

\begin{eqnarray*}
\gamma_{ij}^{(b)} & = & \frac{\alpha_{MLT}u'}{\gamma}\beta^{2}K\left(\beta\right)g_{n}\varepsilon_{inj},
\end{eqnarray*}
where $\alpha_{MLT}$ is the mixing-length parameter, $\gamma$ is
the adiabatic exponent, and $\mathbf{g}$ is the unit vector in the
radial direction. The quenching of the magnetic buoyancy (see \citealt{kp93})
is defined by function $K\left(\beta\right)$: 
\[
K\left(\beta\right)=\frac{1}{16\beta^{4}}\left(\frac{\left(\beta^{2}-3\right)}{\beta}\arctan\left(\beta\right)+\frac{\left(\beta^{2}+3\right)}{\left(\beta^{2}+1\right)}\right).
\]
A detailed formulation has been published by \citet{pi15M}. The distribution
of the turbulent parameters, e.g, such as the typical convective turn-over
time, $\tau_{c}$, the mixing length, $\ell$, the RMS convective
velocity, $u'$, the mean density, $\bar{\rho}$ and its gradient
$\mathbf{\boldsymbol{\Lambda}}^{(\rho)}=\boldsymbol{\nabla}\log\overline{\rho}$
are computed using the mixing-length model of the solar convection
zone of \citet{stix:02}. In particular, it uses the mixing length
$\ell=\alpha_{{\rm MLT}}\left|\Lambda^{(p)}\right|^{-1}$, where $\Lambda{}^{(p)}={\nabla}\log\overline{p}$,
is the inverse pressure height, and $\alpha_{{\rm MLT}}=2$. The turbulent
diffusivity profile is given in the form $\eta_{T}=C_{\eta}{\displaystyle \frac{u'^{2}\tau_{c}}{3f_{ov}\left(r\right)}}$,
where function $f_{ov}(r)=1+\exp\left(50\left(r_{ov}-r\right)\right)$,
controls quenching of the turbulent effects near the bottom of the
convection zone, $r_{ov}=0.725R_{\odot}$ . The parameter $C_{\eta}$,
(in the range $0<C_{\eta}<1$) is a free parameter that controls the
efficiency of large-scale magnetic field mixing by turbulence. It
is used to adjust the period of the dynamo cycle. We use the same
model of the convection zone for all the rotational periods considered
in this paper.

At the bottom of the convection zone we apply a perfectly conducting
boundary condition. At the top of the convection zone the poloidal
field is smoothly matched to the external potential field, and the
toroidal field is allowed to penetrated to the surface: 
\begin{eqnarray}
\delta\frac{\eta_{T}}{r_{e}}B+\left(1-\delta\right)\mathcal{E}_{\theta} & = & 0,\label{eq:tor-vac}
\end{eqnarray}
where $\delta=0.99$ \citep{1992AA256371M,pk11apjl}. The numerical
integration is carried out in latitude from the pole to pole, and
in radius, from $r_{b}=0.715R_{\odot}$ to $r_{e}=0.99R_{\odot}$.
The numerical scheme employs the pseudo-spectral approach for the
numerical integration in latitude and the second-order finite differences
in radius.

\begin{table}
\begin{tabular}{>{\raggedright}p{2cm}>{\raggedright}p{1.5cm}>{\centering}p{3cm}>{\centering}p{2.8cm}>{\centering}p{2.8cm}>{\centering}p{2.8cm}}
\hline 
Runs  & Quenching  & Dynamo Period, {[}YR{]}  & Flux,

10$^{24}${[}MX{]}  & $\overline{\left|B^{\left(p\right)}\right|}$,

{[}G{]}  & Polar, $\left|B_{r}\right|$,

{[}G{]}\tabularnewline
\hline 
M0  & AQ  & 13.2, 14, 17.6,

21, 21.4, 23  & 2.2, 6.4, 11.4,

14.8, 18.3, 20.8  & 10, 12, 15,

23, 25, 27  & 11, 20, 25,

29, 30, 31\tabularnewline
\hline 
M1  & AQ, BQ  & 11.1, 8.8, 6.8,

5, 4.8, 4.6  & 0.7, 1.2, 1.5,

1.7, 1.8, 1.9  & 3, 5.5, 8,

10, 7.5, 6.8  & 2, 4, 4.3,

3.8, 3.6, 3.4\tabularnewline
\hline 
M2 , $R_{m}=10^{6}$ & MHQ  & 12.7, 10.8, 8.1,

6.6, 6.2, 5.2(5.9)  & 0.4, 1., 1.9,

2.7, 3, 3.6  & 1.5, 2.7, 3.1,

3.1, 4.3, 7.5  & 1.5, 2.5, 2.,

3.5, 2.8, 7.8\tabularnewline
\hline 
M3  & MHQ, BQ ,AQ & 11.6, 9.9,7.2,

5.8, 5.4, 5  & 0.4, 0.9, 1.3,

1.8, 1.9, 2.1  & 2.3, 2.5, 3.4,

3.1, 4.5, 7 & 2.2, 2.3, 1.9,

2.5, 3.2, 4\tabularnewline
\hline 
M3A, $C_{\alpha}=0.03$ & -/- & 12.4, 11, 8.4,

6.5, 6.1, 5.5 & 0.3, 0.6, 1.1,

1.5, 1.7, 1.9 & 1.3, 2.4, 3.2,

3.5, 3.6, 4.5 & 1.3, 2.1, 1.7,

1.7, 2.1, 2.7\tabularnewline
\hline 
M3B, $C_{\alpha}=0.06$ & -/- & 9.5, 7.6, 5.7,

4.7, 4.5, 4.0 & 0.8, 1.3, 1.9,

2.3, 2.5, 2.8  & 3.9, 4.7, 7.2,

11.3, 13.7, 16 & 3.2, 2.9, 4.6,

20, 28, 44\tabularnewline
\hline 
\end{tabular}\caption{Results of the model runs for the set of six rotational periods $P_{rot}=\left[29.4,25,20,16.7,15.6,14.3\right]$
days from left to right in each cell. Four types of the dynamo models
with different non-linear quenching mechanisms (second column): the
algebraic quenching (AQ), the magnetic buoyancy quenching (BQ) and,
the magnetic helicity quenching (MHQ). The other columns show the
dynamo period, the total unsigned magnetic flux, the mean poloidal
magnetic field strength at the surface, $\overline{\left|B^{\left(p\right)}\right|}$,
the strength of the radial polar field, $\left|B_{r}\right|$ . }
\end{table}

\section{Results}

\begin{figure*}
\includegraphics[width=0.8\paperwidth]{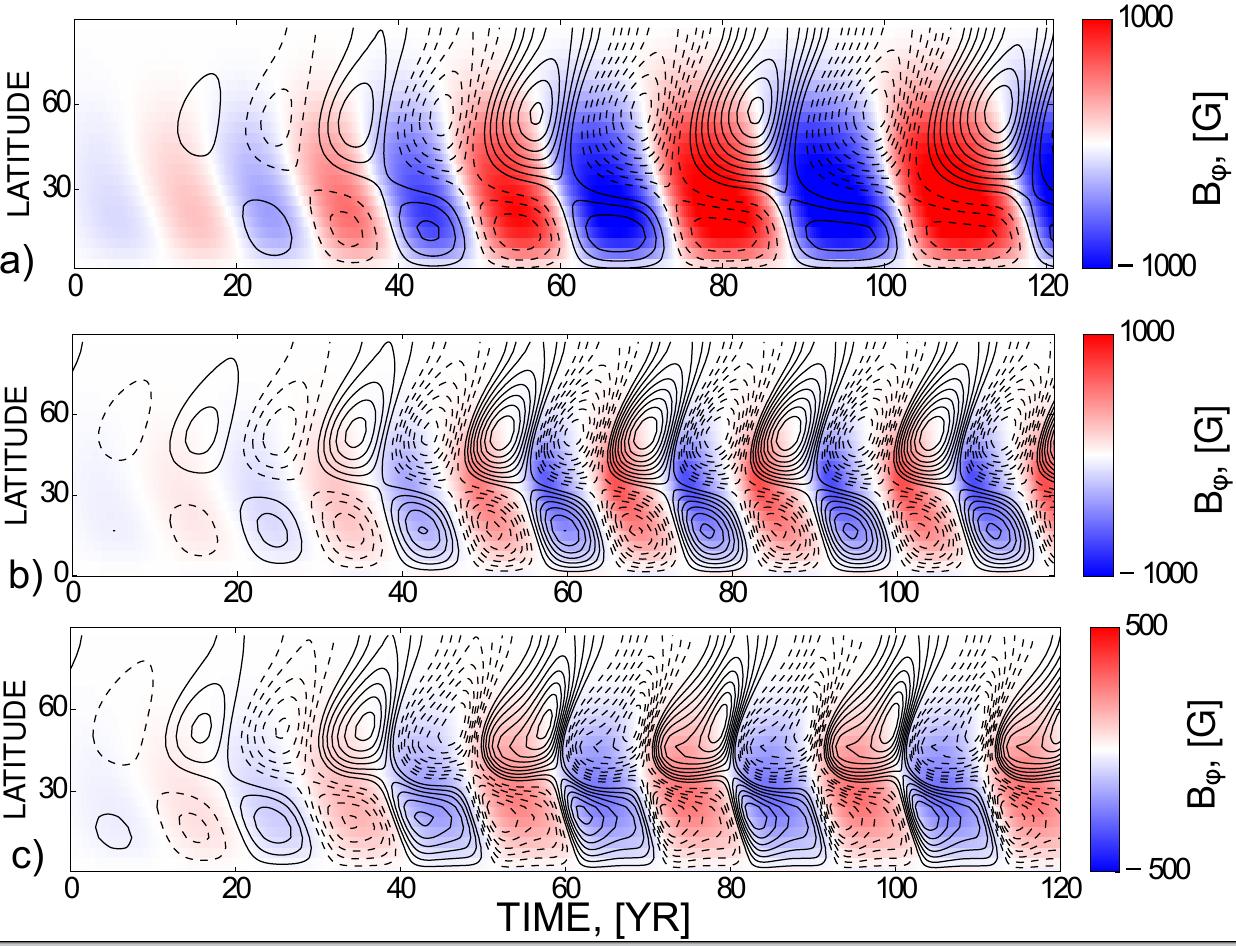}\caption{Time-latitude diagrams for the radial magnetic field at the surface
(contours) and the toroidal field in the subsurface layer (background
image) for the rotational period of 25 days: a) model M0; b) model
M1; c) model M2 (Table 1). The strength of the radial field varies
in range of $\pm$20G in model M0, $\pm5$G in model M1 and $\pm3$G
in model M2.}
\end{figure*}

The free parameters, $C_{\alpha}$, $C_{\eta}$ and $R_{m}$ are used
to calibrate the model for the best possible agreement with the solar-cycle
observations for the solar rotation period. In this study we use $C_{\alpha}=0.04$,
$C_{\eta}=0.05$ and $R_{m}=10^{4}$. In model M2 (Table 1) we use
$R_{m}=10^{6}$, which provides a better conservation of magnetic
helicity, which is the primary non-linear effect in this case. The
profile of the rotation law is taken from the helioseismology inversion
of \citet{Howe2011JPh}. It is fixed, as well (see, \citealt{pi15M}).
However, the radial profile of the Coriolis number $\Omega^{*}=4\pi{\displaystyle \frac{\tau_{c}}{P_{rot}}}$
varies with the rotational period. In our runs, we employ the following
set of rotational periods: 
\begin{equation}
P_{rot}=\left[29.4,25,20,16.7,15.6,14.3\right]\label{eq:pset}
\end{equation}
For this set we compute the dynamo models (Eqs(\ref{eq:dyn},\ref{eq:helcon-1}))
taking into account the different saturation mechanisms, as specified
in Table 1. The models which are listed in Table 1 employ the $\alpha$
effect parameter, $C_{\alpha}=0.04$. This value is of twenty percents
above the threshold for the $P_{rot}=25$ days. Table 1 also shows
results for the dynamo cycle period, the magnitude of the unsigned
magnetic flux generated in the whole convection zone, the mean strength
of the poloidal magnetic field at the surface, $\overline{\left|B^{\left(p\right)}\right|}$
, (which we will use as a proxy of the line-of-sight magnetic field
strength), and the strength of the radial polar magnetic field, $\left|B_{r}\right|$,
during the cycle minimum. The dynamo models are calculated untill
they reach a stationary phase. In the study we consider the magnetic
field geometry antisymmetric relative to the equator. This is done
via imposing a weak seed poloidal magnetic field of the dipole symmetry
in the initial conditions. Model M3 was explored for a set of the
$\alpha$ effect parameter, $C_{\alpha}=[0.03,0.04,0.06]$. Also,
for model M2 we made additional runs for the rotational periods 16.7
and 15.6 days using $C_{\alpha}=0.06$ and $C_{\alpha}=0.05$ respectively
to study the behavior of the long-term variations in the dynamo process
for different rotational periods.

\begin{figure}
\includegraphics[width=0.8\linewidth]{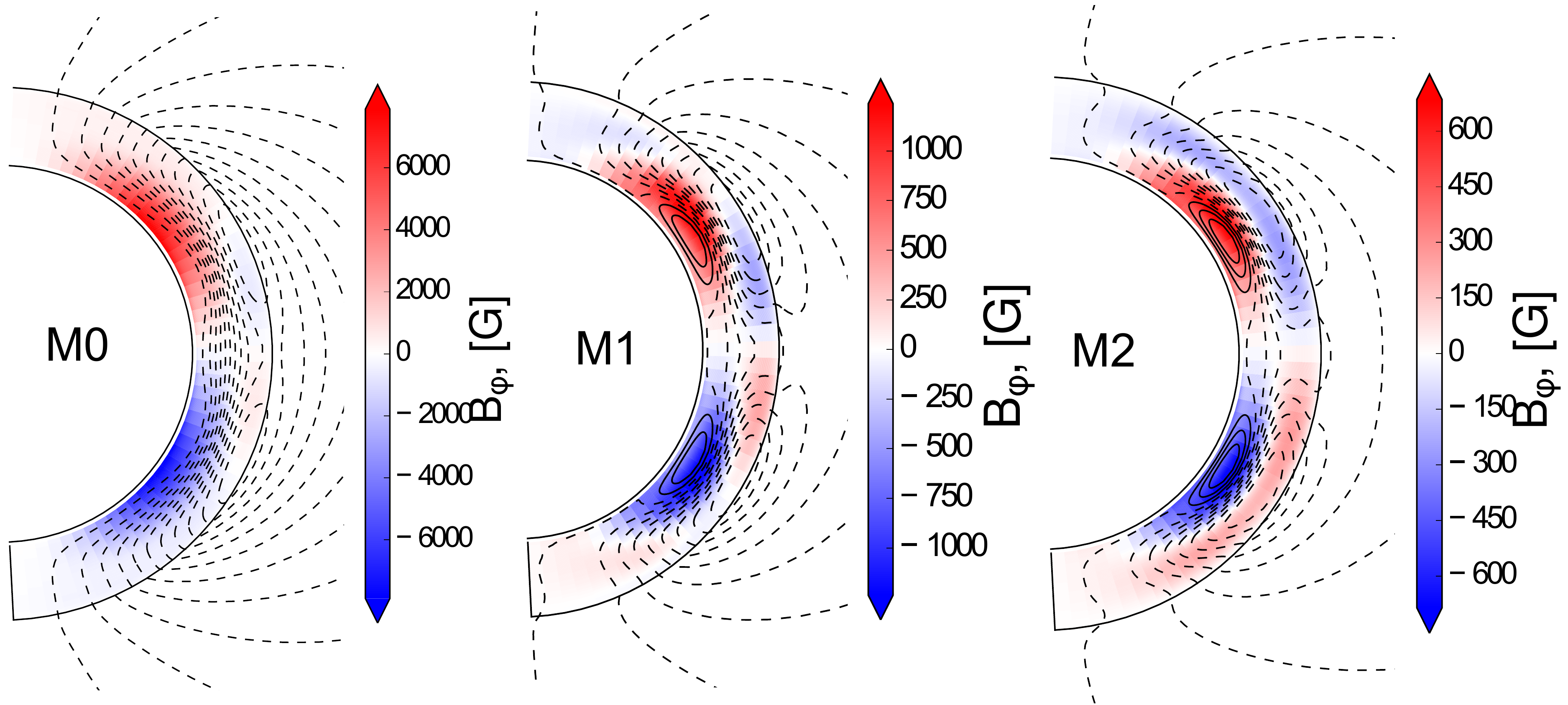}\caption{Snapshots of magnetic field distributions in the minimum of the magnetic
cycle for models M0, M1 and M2 (Table 1) for the samerotational period
25 day, the color background image shows the strength of the toroidal
magnetic field and contours show streamlines of the poloidal magnetic
field. The strength of the poloidal field is given in Table 1.}
\end{figure}

\begin{figure*}
\includegraphics[width=0.8\paperwidth]{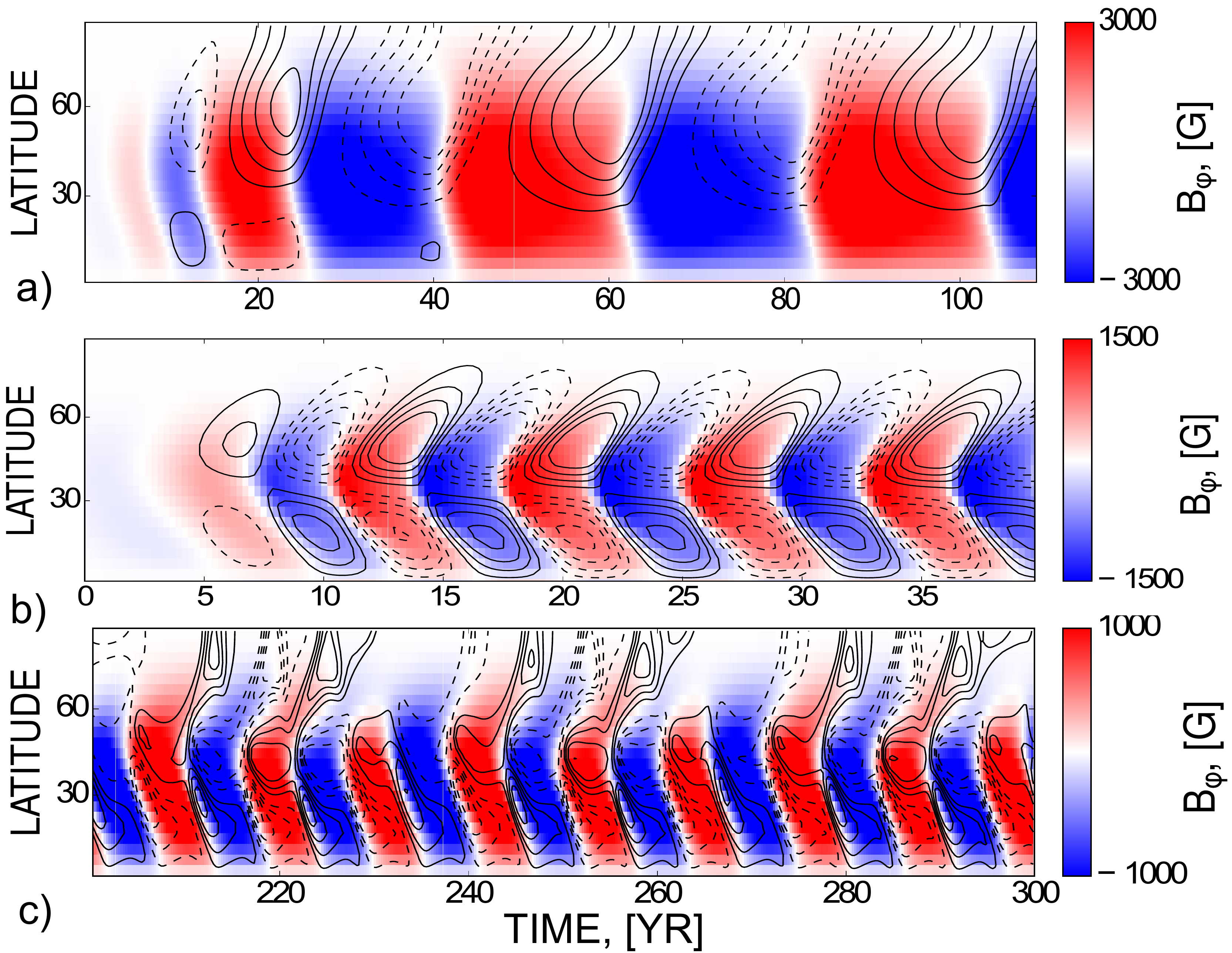}\caption{The same as Fig.1 for the rotational period 14 days. Model M2 has
a much longer relaxation time. The radial magnetic field varies in
range $\pm25$G in model M0, $\pm5$G in M1 and $\pm15$G in M2.}
\end{figure*}

The effect of the quenching mechanisms on the time-latitude evolution
of the toroidal and poloidal magnetic field is illustrated in Figures
1, 2 and 3, where we show results for the rotational periods of 25
and 16.7 days for models M0, M1 and M2. For the rotation period of
25 days all models produce the solar-like time-latitude magnetic butterfly
diagrams. Model M0 has the strongest large-scale magnetic field, and
it has the longest dynamo period among all the runs (see Table 1).
It is also seen that the dynamo period in model M0 increases when
the dynamo approaches the stationary state. This seems to be due to
the strong concentration of the dynamo wave near the bottom of the
convection zone in model M0. This is illustrated by snapshots of the
magnetic fields distributions in Figure 2.

Increasing the rotation rate results in an increase of the dynamo
period in model M0 because the toroidal field is amplified near the
bottom of the convection zone. Simultaneously, the $\alpha$ effect
is suppressed there. This results in a spatial separation of the dynamo
generation by the $\alpha$ and $\Omega$ effects. \citet{deinz74}
had shown that this causes an increase of the dynamo period. Note,
that in the flux-transport models the $\alpha$ and $\Omega$ effects
are spatially separated by design of the models\textbf{.} Models M1,
M2 and M3 show a decrease of the dynamo period when the period of
rotation decreases, as they preserve a distributed character of the
dynamo process. 

Increasing non-linearity in the dynamo process can increase the complexity
of the dynamo evolution. This happens in model M2 for the rotational
period of 14 days. The model shows a long-term modulation with a period
of about 40 years while having two primary dynamo periods of 5.3 and
5.9 years. The modulation disappears when the kinetic helicity parameter
$C_{\alpha}$ increases. Similar long-term modulations were found
for the rotational periods 16.7 and 15.6 days with $C_{\alpha}=0.06$
and $C_{\alpha}=0.05$ respectively. The long-term modulation in axisymmetric
dynamo models with the dynamical quenching of the $\alpha$ effect
was demonstrated earlier in the number of papers (see, e.g., \citealt{cov98,kitiash09}).
Model M2 illustrates an interesting possibility when the surface radial
magnetic field almost disappears during the maximum of the grand cycle
of the toroidal magnetic field. A situation when the toroidal field
dominates on the stellar surface is also observed in stellar activity,
but for a non-axisymmetric field, and for the faster rotating star,
see, e.g., the review by \citet{2009ARAA_donat}.

\begin{figure}
\includegraphics[width=0.6\columnwidth]{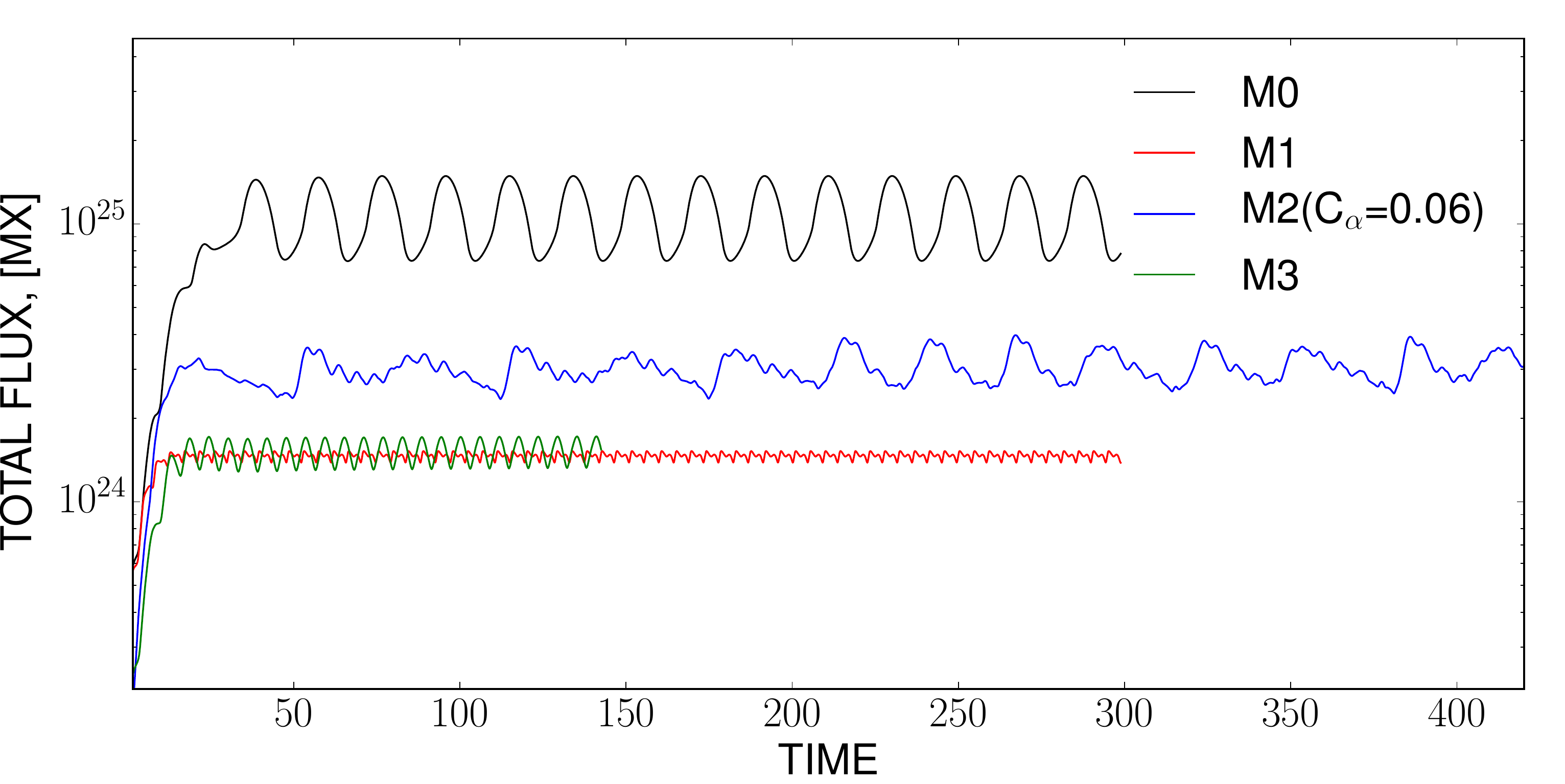}

\caption{\label{fig:flux}Variations of the total unsigned flux of toroidal
field in the convection zone for the star rotating with period of
16 days. Model M2 has long-term modulation of the magnetic activity
for $C_{\alpha}=0.06$.}

\end{figure}

Variations of the total unsigned flux of the toroidal field in the
convection zone for the star rotating with the period of 16 days are
shown in Figure \ref{fig:flux}. It is seen that the non-linear mechanism
involved in the dynamo affects the amplitude of the dynamo generated
flux, the range and period of the variations. Model M2 has a long-term
modulation of the magnetic activity for $C_{\alpha}=0.06$. The same
effect is found for model M2 with the rotational period of 14 days
and $C_{\alpha}=0.04$ (see Figure 3c). It is interesting that individual
cycles are not always well recognized in the long-term modulations
of the flux. 

\begin{figure*}
\includegraphics[width=0.76\paperwidth]{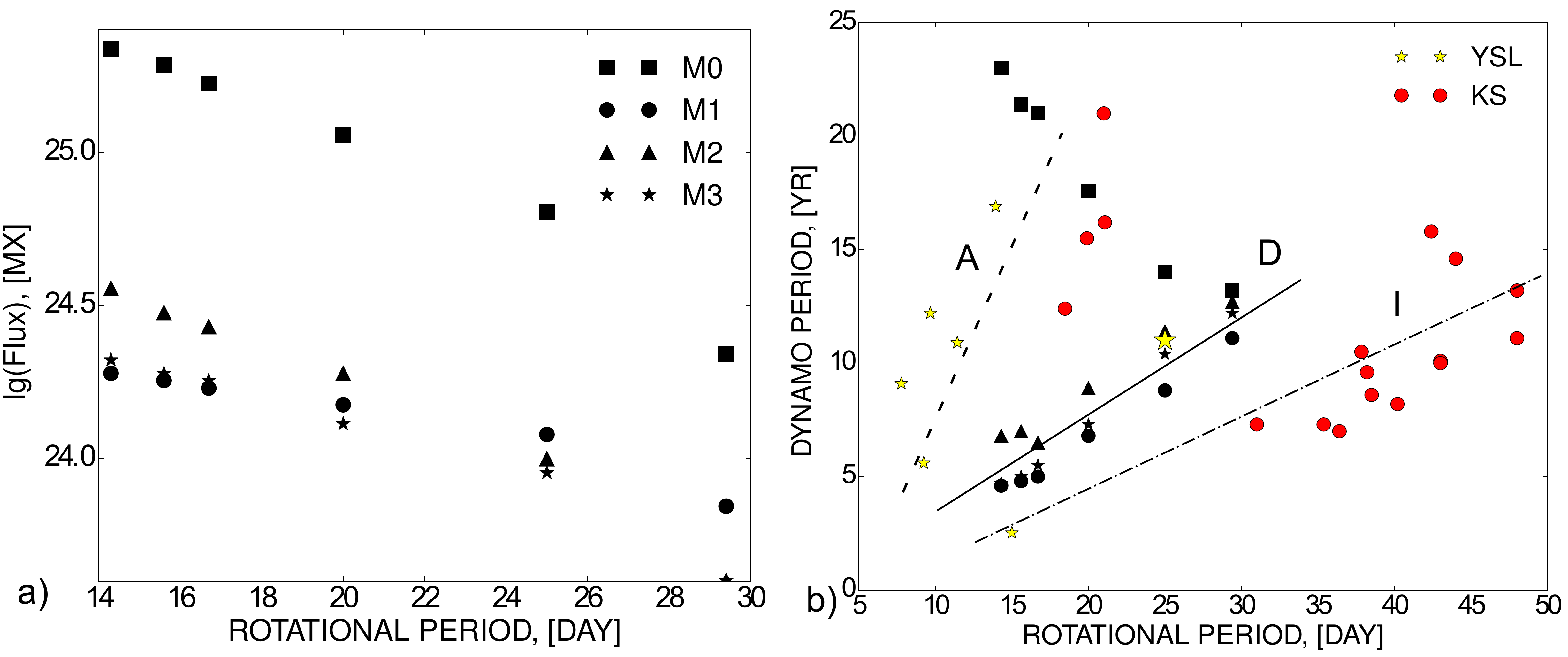}\caption{\label{fig-flux-pp}a) Dependence of the amplitude of dynamo generated
magnetic flux in the convection zone on the period of rotation; b)
the dynamo period vs the rotational period in the models (Table 1)
and in sample of young solar like stars, denoted as YSL (the sample
of G-stars from \citealp{2007ApJ_bohm}), KS denotes the sample of
K-stars from the same paper, the Sun is marked by big yellow symbol
star. The ``A'' (dashed line) stands for the active branch, ``I''
(dash-dot line) - the inactive one and ``D'' (solid line) - the
branch produced by our models M1 and M2.}
\end{figure*}

Figure \ref{fig-flux-pp}a shows the dependence of the dynamo generated
unsigned magnetic flux in the convection zone on the period of rotation.
It shows an increase of the generated flux with an increase of the
rotation rate. It is seen that the magnetic helicity conservation
produces the strongest quenching of the dynamo process among all the
non-linear mechanisms.

Thus, variations of the dynamo period with the rotational period depend
on the dynamo saturation mechanism. Figure 4b illustrates our results
together with the observational results shown earlier by \citet{2007ApJ_bohm}.
The algebraic quenching of the $\alpha$ effect is a simple anzatz
most widely used in various dynamo models. However, this non-linear
effect results in an increase of the dynamo period with increasing
rotation rate. This tendency disagrees with the observations. The
opposite trend is demonstrated by models M1, M2 and M3. It qualitatively
agrees with the observations that show a complicated behavior of the
stellar activity periods, indicating the existence of two populations
of magnetically ``active'' and ``inactive'' stars \citep{1999ApJ_sa_br,2007ApJ_bohm,2011IAUS_saar}.
\begin{figure}
\includegraphics[width=0.7\columnwidth]{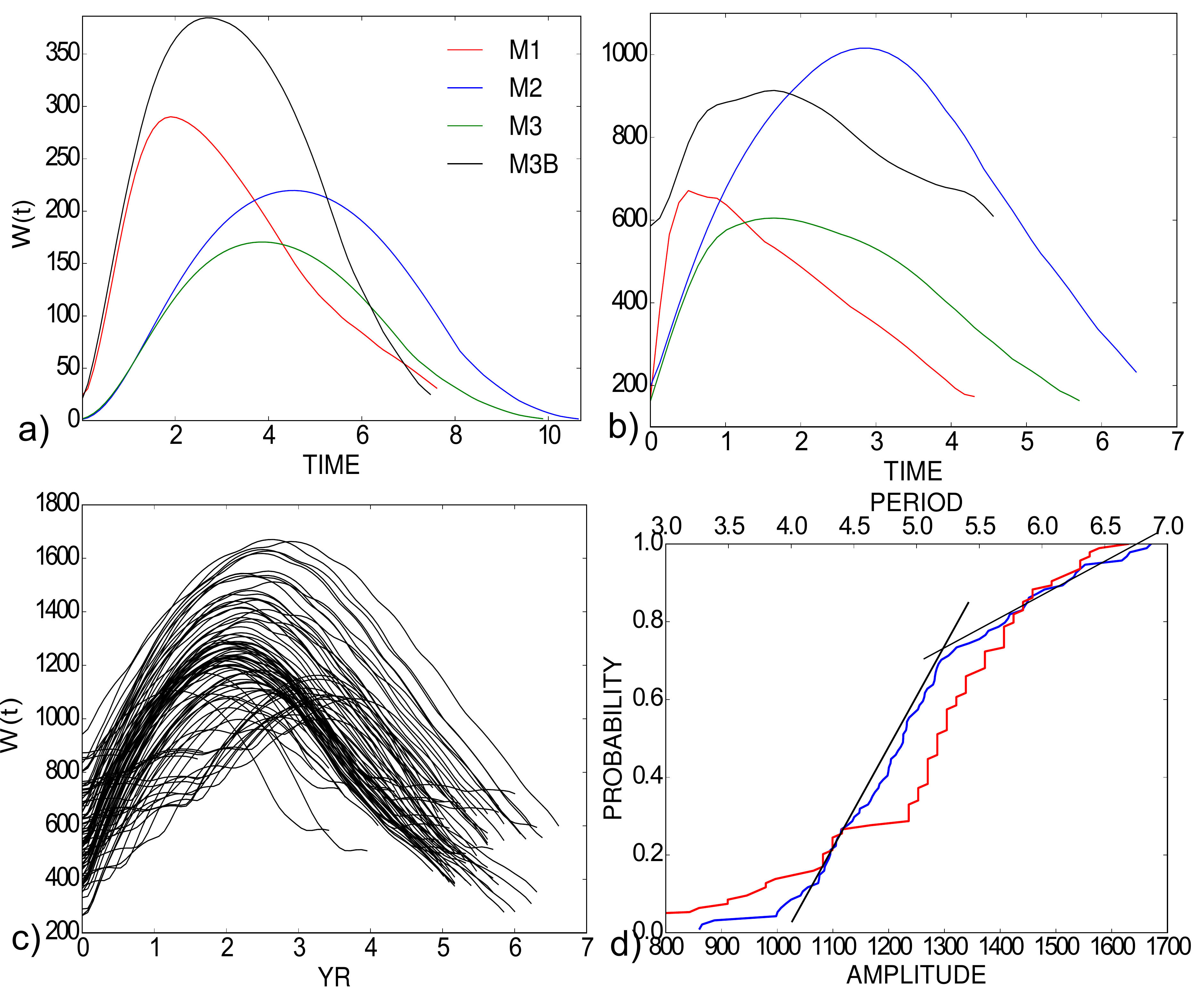}

\caption{\label{fig:Wolf}The Wolf number for the individual cycles: a) the
rotational period of 25 days; b) the same for the rotational periods
of 16 days; c) the set of cycles for the model M2 with long-term modulation,
the rotational period 16 days and $C_{\alpha}=0.06$; d) Probability
of the cycle amplitude(blue) and period (red) in the long-term modulation
of magnetic activity. Two intervals with the different uniform probability
distributions of the cycle amplitude are marked by the linear trends.}

\end{figure}

It is interesting to estimate how the magnetic activity changes parameters
of spotness during the magnetic cycle. For this purpose we employ
a simple relation between the Wolf susnspot number $W\left(t\right)$
and the amplitude of the toroidal magnetic field in the subsurface
layer $B_{max}$, 
\begin{equation}
W={\displaystyle B_{max}\exp\left(-\frac{B_{0}}{B_{max}}\right)},\label{eq:Wolf}
\end{equation}
where $B_{0}=1000$G \citep{pipea2012AA}. The parameter $B_{0}$
is slightly different to that one used in the above cited paper, because
here the $B_{max}$ value is estimated from the layer in the upper
convection zone where the maximum of the dynamo wave is located. Thus,
the depth of the $B_{max}$ is not fixed and it varies with time.
This provides a smoother profile of $W(t)$ for the fast rotating
stars compared to the previous definition. Examples of the Wolf number
profiles for different models are shown in Figure \ref{fig:Wolf}.
The cycle of the model M0 is substantially larger than in other models
and it is not shown. It is seen that magnetic buoyancy is responsible
for the most asymmetric profile of $W\left(t\right)$. Also, it is
found that the rise time decreases when the cycle amplitude increases.
This effect works in both cases for the decrease of the rotational
period and for the increase of the $\alpha$-effect amplitude. Similar
phenomenon is observed in solar activity \citep{vetal86}.

All models listed in Table 1, except model M2 with the rotational
period 14 days, reach a stationary state, which is characterized by
a non-linear oscillation of some fixed period. For the rotational
periods of 14, 15, and 16 days we find long-term modulations in model
M2 with $C_{\alpha}=\left[0.04,0.05,0.06\right]$ respectively. In
these cases the basic dynamo period has no fixed value. For each time
series of the theoretical Wolf number parameter $W$, we extract the
individual cycles and compute the cycle parameters, i.e., the magnitude
of the maximum, the period, the rise time and the decay time. We do
this in the same way as previously in \citet{pk11}. Figure \ref{fig:Wolf}c
shows variations of $W$ for the individual cycle profiles. These
profiles were extracted from the time series of model M2 with the
rotational periods of 16 days, and $C_{\alpha}=0.06$. It is seen
that the stronger cycles have longer period in this set. The amplitude
of the period variations is about 3 years. The amplitude of $W$ varies
about twice of the minimal amplitude. The probability distribution
functions are constructed as follows. We sort the amplitudes of $W$
in the increasing order and compute the cumulative probability distribution:
\begin{equation}
CDF=\frac{\sum_{k=1}^{i}k}{N},\label{cdf}
\end{equation}
where, $i$ is the order number (after sorting the set in the increasing
order), and $N$ is the total number of instances. In the set shown
in Figure \ref{fig:Wolf}c $N=92$. Equation (\ref{cdf}) approximates
the probability for the amplitude of $W$ to have values in the interval
between $W_{min}$ and $W_{i}$. The accuracy of the approximation
improves as $N\rightarrow\infty$. The same procedure was repeated
for the period parameter. The result is shown in Figure \ref{fig:Wolf}d
(red curve). From this analysis we conclude that the amplitude of
$W$ demonstrates the existence of two populations of the stellar
cycles, which can be quantified as the ``strong'' and ``weak''
cycles. The two populations are not well recognized from the distribution
of the dynamo period. Other examples of application of Eq(\ref{cdf})
to analysis of the dynamo cycles were presented by \citet{ps11}.

\begin{figure}
\includegraphics[width=0.95\columnwidth]{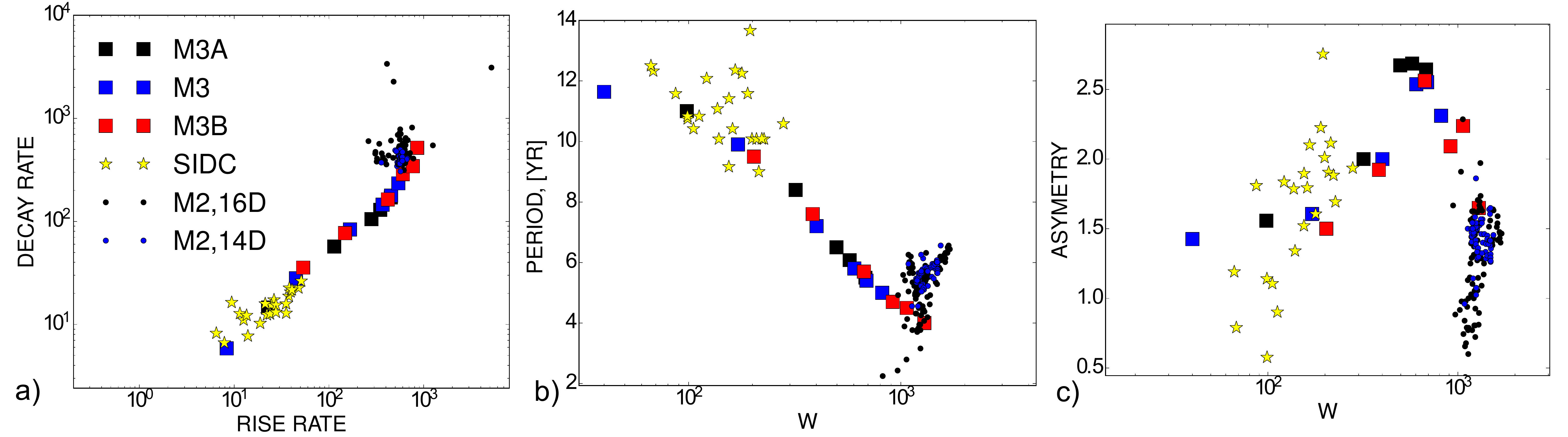}

\caption{\label{wald}Theoretical Waldmeier's relations in stellar cycles.
The ``SIDC'' is related to the sunspot database \citep{sidc}}

\end{figure}

It is interesting to look how the Waldmeier relations, well-known
for the sunspot cycles, and the asymmetry of stellar cycles can change
with the rotational period, and the magnitude of the $\alpha$ effect.
For this we consider model M2 and M3. For model M2 we use only runs
with the long-term modulations. Similarly to \citet{pk11} we determine
the mean rise and decay rate for each individual cycle in each simulatied
time series of $W$. The asymmetry is determined as a ratio between
the decay and rise rates. Figure \ref{wald} shows our results. Firstly,
it is seen that this ratio seems to be constant for the set of the
periodic dynamo models (without long-term modulations). The decay
rates of the dynamo cycles vary consistently with the rise rate, with
the increase of the alpha effect parameter $C_{\alpha}$ and the rotation
rate. The theoretical trend is consistent with observations of the
solar activity cycles. Here we used the data set provided by \citet{sidc}.
However, the time series for model M2 for the rotational periods of
14 and 16 days, which has the long-term modulation of the magnetic
cycle are out of the trend. Figure \ref{wald}b shows that these models
have a correlation between the cycle amplitude and its period. That
contradicts to the general trend of the periodic models and the solar
observations as well. It is interesting to note that populations of
the weak cycles in the long-term modulations follows the anti-correlation
between the amplitude and period of cycle. The effect is not well
recognized in Fig.\ref{wald}b because the scatter of the cycle parameters
is relatively small. We find that this scatter is decreased with the
increasing rotation rate. Also The periodic models show that asymmetry
increases with the increase of the cycle magnitude (Fig.\ref{wald}c).
The trend changes in stars with the high rotation rate and strong$\alpha$-effect.
This is likely because our parameterization of the cycle asymmetry
is too simple and it does not take into account the possible complicated
form of $W$. For example, Fig.6b shows the profile of $W$ , in which
a sharp rise of activity ends by a plateau, and the maximum is not
well defined. The same arguments should be taken into account in analysis
of the sunspot cycles. The models with the nonlinear long-term modulations
do not show a dependence of the asymmetry on the cycle magnitude.

\begin{figure}
\includegraphics[width=0.5\paperwidth]{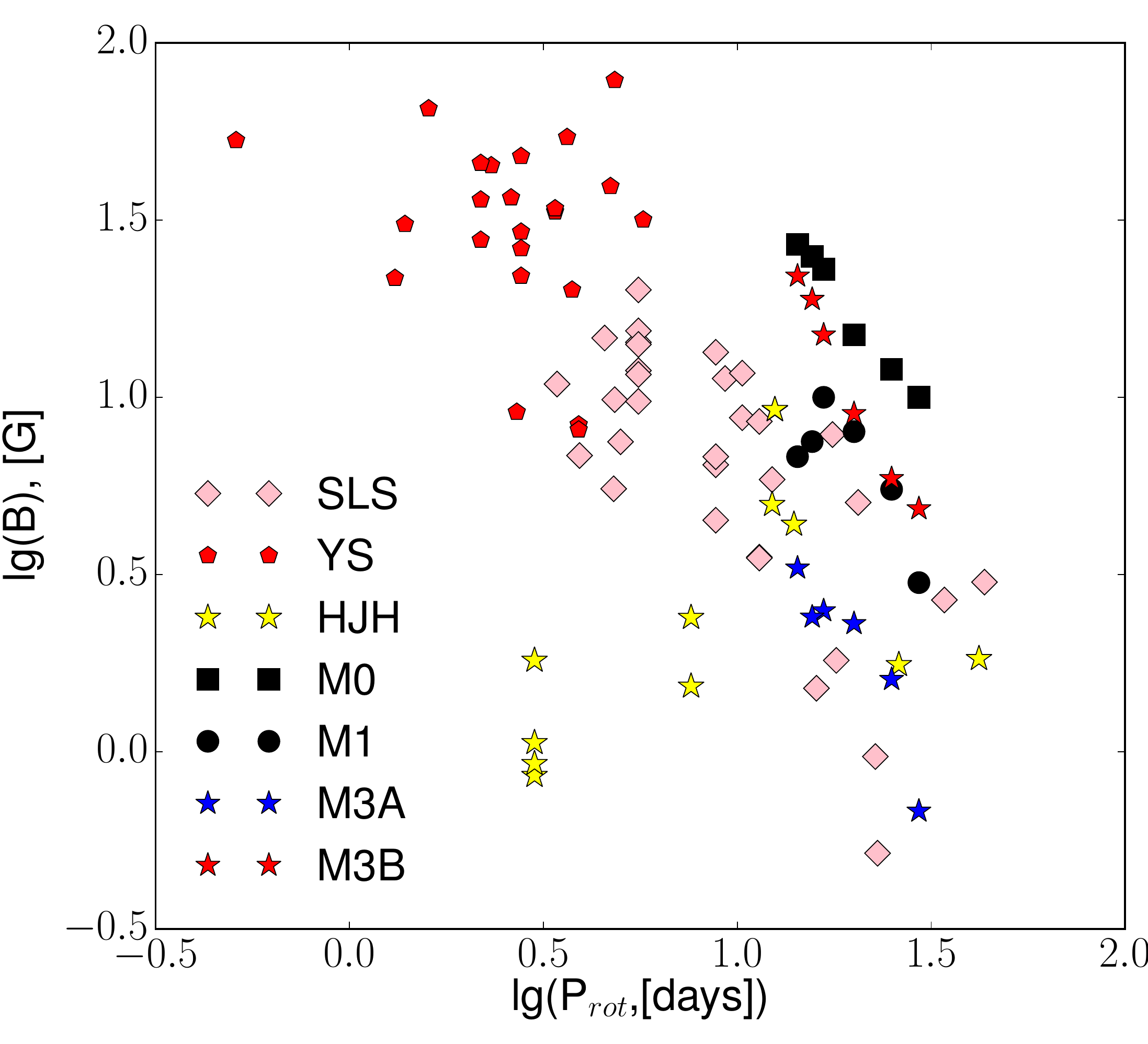}

\caption{Comparison of our dynamo models with results of survey of \citet{vid14MNR}:
the mean line-of-sight magnetic field on the surface vs the period
of rotation, the SLS symbols show the solar-like stars, the YS - the
young Suns, the HJH - stars host of hot Jupiters, notations of the
model results are the same as in Figure 4. }
\end{figure}

Figure 5 shows a comparison of our dynamo models with the results
of survey of \citet{vid14MNR} for the absolute value of the mean
line-of-sight magnetic field on the surface versus the period of rotation.
We see that all the models qualitatively agree with the results of
observations within the investigated range of the rotational periods.
Model M0 shows a stronger magnetic field than it is found in the observations.
This indicates on the presence of additional dynamo saturation mechanisms,
like those implemented in models M1 and M2, i.e., the nonlinear quenching
due to the magnetic buoyancy or the magnetic helicity. As noted in
Introduction, for rapidly rotating stars the magnetic feedback on
the differential rotation and convection has to be taken into account
as well. Models M3A and M3B show different results for the different
magnitude of the $\alpha$-effect parameter, $C_{\alpha}$. We see
that the observed scatter of the magnetic field strength can be explained
by variations of $C_{\alpha}$, varing by a factor of two above the
dynamo instability threshold. From the solar dynamo models results
we know that a similar effect can be produced by a small fluctuations
of $\alpha$ on the time scale comparable with the length of the magnetic
cycle \citep{moss-sok08}. 

\begin{figure}
\includegraphics[width=0.5\linewidth]{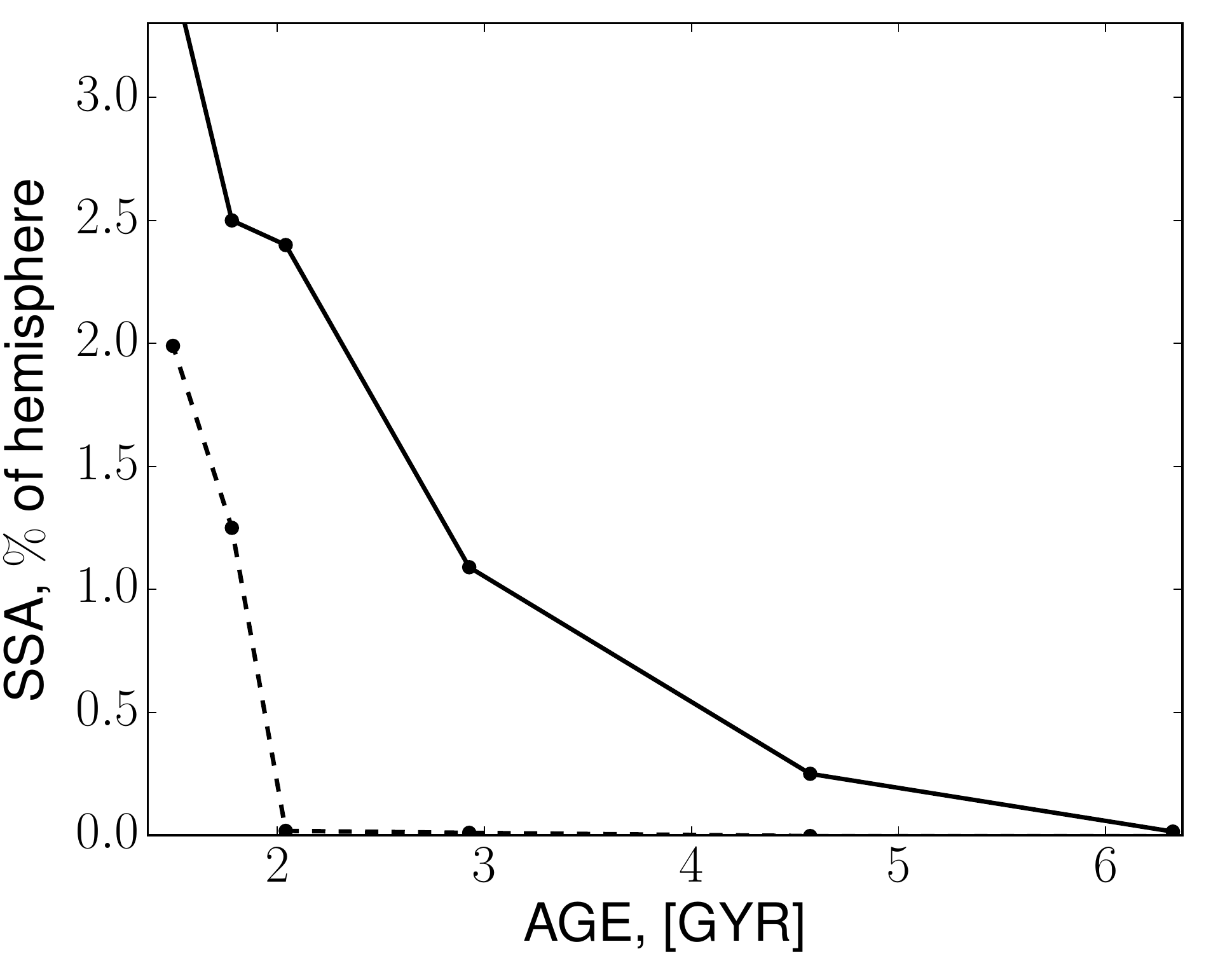}

\caption{The theoretical sunspot area calculated for model M2, solid line shows
the maximum value and the dashed line shows the minimum.}
\end{figure}

The investigated sample of rotational periods can be related to different
ages of the sun-like stars. Taking into account the slowdown of rotation
with time as $t^{-1/2}$, where $t$ is the star's age \citep{skum72},
we deduce that age of stars in our sample can be found from the relation:
${\displaystyle t_{*}=t_{\odot}\left(\frac{P_{*}}{P_{\odot}}\right)^{2}}$
, where $P_{\odot}=25$ days and $t_{\odot}=4.61$ Gyr are parameters
of the modern Sun. Thus, the investigated sample of the rotational
periods covers the range of ages from 1.5 to 6.2 Gyr. However, on
this time scale changes of the convection zone structure have to be
taken into account. The percentage of disk occupied by sunspots is
estimated from the empirical correlation $SSA\approx16.7W/10^{4}$
\citep{vetal86}. Figure 6 shows our estimation for amplitude of $SSA$
at the maximum and minimum (dashed curve) of magnetic cycle with the
age for model M2. It shows that the younger stars are significantly
more spotty, This is a rather naive estimate which does not take into
account mechanisms of the sunspot formations, and also changes of
the convection zone and the differential rotation with the stellar
age.

\section{Discussion and conclusions}

We have studied the nonlinear magnetic feedback on the turbulent generation
of the large-scale magnetic fields by a solar-type dynamo for the
range of rotation periods from 14 to 30 days. Our models take into
account the non-linear $\alpha$ effect, the balance of magnetic helicity
density, and escape of magnetic field from the dynamo region due to
magnetic buoyancy. We did not consider the possible magnetic feedback
on the differential rotation and changes in the convection zone stucture. 

Currently, one of the most important observational tests for dynamo
models of solar-like stars is the relationship between the magnetic
cycle periods and the period of stellar rotation. Earlier, \citet{1984ApJ...287..769N},
also \citet{oss97} and \citet{tob98} had suggested that despite
a seemingly linear character of the connection between the rotational
period and the dynamo period in stellar magnetic activity, the relation
is not trivial because the large-scale dynamo goes to a non-linear
regime when the rotation rate increases. The revealed multiple populations
of magnetic activity in late-type stars \citep{1999ApJ_sa_br} showed
that the theoretical interpretation of stellar magnetic activity can
be complicated. It was suggested that the ``active'' population
of stars operates the dynamo concentrated near the bottom of the convection
zone, and the ``inactive'' population that operates \citep{1999ApJ_sa_br,2007ApJ_bohm}
a distributed dynamo. In most cases, the models of the first type
show an increase of the dynamo period with increasing of rotation
rate \citep{brun10AA,2014ApJ...791...59K,pi15M} and are not consistent
with observations. Our results show that the observed correlation
between the period of rotation and the dynamo period can be explained
by the distributed dynamo models that includes the dynamical magnetic
feedback on the turbulent generation either due to the magnetic buoyancy
or the magnetic helicity. Note that the magnetic buoyancy effect which
is employed in the paper is related to buoyancy of large-scale magnetic
fields and not to the so-called flux-tube buoyancy \citep{park}.
\citet{kp93} discussed the issue in details. 

In our set of dynamo models we did not find a satisfactory explanation
of the observed multiple populations in terms of the stellar activity
cycles periods. A number of possibile hypothesises remains to be explored.
For example, our study is limited to magnetic field configurations
that are anti-symmetric relative to the equator. The magnetic parity
breaking is often considered as a source of non-linear long-term modulations
of solar activity \citep{1976SvA20227I,bran89,sok1994AA,2016MNRAS.456.2654W}.
This may be the case for the stellar activity as well. The correlation
between the period of rotation and the dynamo period could be affected
by changes of the convection zone characteristics and changes of the
differential rotation profile \citep{kit-r11}. Models with long-term
modulations and dynamical quenching of the $\alpha$-effect by the
magnetic helicity conservation show two populations of the activity
cycles. The population of the weak cycles tends to follow standard
the Waldmeir's rules. But these do not hold for the whole time series. 

\citet{2009ARAA_donat} presentented a diagram of stellar magnetic
activity for the period of rotation versus the stellar mass diagram
of magnetic activity, which shows that fast rotating solar mass stars
have non-axisymmetric magnetic fields. It is not clear if the non-axisymmetric
field on the fast rotating solar analogs is generated by the dynamo
instability or results from emergence of magnetic field on the surface.
The non-axisymmetric fields is rarely discussed in the solar type
dynamo. It presumes that the differential rotation suppresses the
non-axisymmetric dynamo \citep{rad86AN}. However, for the $\alpha\Omega$
dynamos, the critical threshold for generation of non-axisymmetric
magnetic fields is only about a factor 2 or 3 larger than those for
the axisymmetric dynamo. The interaction of the axisymmetric and non-axisymmetric
modes was never studied for the super-critical regimes which could
be operated in fast rotating stars. Finite non-axisymmetric perturbations
can also affect the axisymmetric dynamo \citep{pk15}. Future studies
should show if the observed multiple populations of magnetic cycles
in stellar activity can be explored by differences between the axisymmetric
and non-axisymmetric dynamo models.

VVP is supported by RFBR grant and the project II.16.3.1 of ISTP SB
RAS. AGK is partially supported by NASA grant NNX14A070G

 
\end{document}